\documentclass[two column, prl]{revtex4}
\usepackage{graphics,graphicx}
\usepackage{ulem}
\begin{document}
\title{Fractional electrical dimensionality in the spin solid phase of artificial honeycomb lattice}
\author{A. Dahal}
\author{B. Summers}
\author{D. K. Singh$^{*}$}
\affiliation{Department of Physics and Astronomy, University of Missouri, Columbia, MO 65211 USA}
\affiliation{$^{*}$email: singhdk@missouri.edu}

\begin{abstract}
Two-dimensional artificial magnetic honeycomb lattice is at the forefront of research on unconventional magnetic materials. Among the many emergent magnetic phases that are predicted to arise as a function of temperature, the low temperature spin solid phase with zero magnetization and entropy is of special importance. Here, we report an interesting perspective to the consequence of spin solid order in an artificial honeycomb lattice of ultra-small connected elements using electrical dimensionality analysis. At low temperature, $T \leq$ 30 K, the system exhibits a very strong insulating characteristic. The electrical dimensionality analysis of the experimental data reveals a fractional dimensionality of $d$ = 0.6(0.04) in the spin solid phase of honeycomb lattice at low temperature. The much smaller electrical dimension in the spin solid phase, perhaps, underscores the strong insulating behavior in this system. Also, the fractional dimensionality in an otherwise two-dimensional system suggests a non-surface-like electrical transport at low temperature in an artificial honeycomb lattice. 
\end{abstract}

\maketitle

The honeycomb lattice structure has generated significant research interest in recent time, primarily motivated by unusual electronic and magnetic properties, as found in Graphene, silicene, MoS$_{2}$ and artificial magnetic honeycomb lattice, respectively.\cite{Sasha,Feng, Ataca,Nisoli} Artificial magnetic honeycomb lattice, initially conceived to explore the statistical properties of spin ice system in two-dimension,\cite{Tanaka} has raised many distinct issues and developed into its own field of study. At the core of it lies the local moment arrangements of 'two-in \& one-out' (or vice-versa) or 'all-in or all-out' configurations on a given vertex of the lattice that give rise to highly degenerate ground state.\cite{Nisoli} Here, 'two-in \& one-out' refers to two moments, aligned along the length of the honeycomb element, point to the vertex and one points away from the vertex, see Fig. 1a. Similarly, the 'all-in' configuration indicates that all moments are pointing to the vertex of the lattice, Fig. 1b. The ease in tunability of the structure allows us to study a plethora of magnetic phenomena, such as spin ice, spin liquid and topological magnetic properties, in a disorder free environment.\cite{Branford} Recent theoretical calculations predict the presence of a rich phase diagram of emergent spin correlated phases in artificial magnetic honeycomb lattice that would be difficult to achieve in a naturally occurring bulk material. It includes the temperature dependent evolution of spin ice order to spin liquid state, followed by an entropy-driven magnetic charge-ordered state as a function of reducing temperature.\cite{Nisoli,Branford} At much lower temperature, the system is predicted to develop a spin solid order, consisting of an alternating arrangements of vortex loops of opposite chiralities.\cite{Moller,Chern} Since the vortex magnetization profile implies zero net magnetic moment, the spin order in a honeycomb lattice is expected to exhibit zero magnetization and entropy state. Detailed understanding of a magnetic material or a particular physical or magnetic state requires the knowledge of electrical properties, in addition to magnetic properties. In this letter, we investigate the effect of spin solid order on the electrical properties in an artificial magnetic honeycomb lattice.

The formation of vortex loops in the spin solid state is expected to have profound impact on the electrical characteristic of the two-dimensional magnetic honeycomb lattice.\cite{Le} However, not much is known about the low temperature electrical properties in this system. Electrical measurements on the recently designed two-dimensional permalloy (Ni$_{0.81}$Fe$_{0.19}$) artificial honeycomb lattice of ultra-small elements reveal strong insulating characteristic at low temperature, T $\leq$ 30 K, when the system is known to manifest a spin solid order.\cite{Brock} Further understanding about the nature of electrical transport is gained by analyzing the electric data at low temperature using electrical dimensionality analysis. In a complete surprise, we find that the system depicts a fractional electronic dimensionality, $d$ = 0.6(0.04), as opposed to the two-dimensional character. It suggests that the electrical transport is a non-surface-like phenomenon, perhaps represents a path-like behavior.

\begin{figure}
\centering
\includegraphics[width=8 cm]{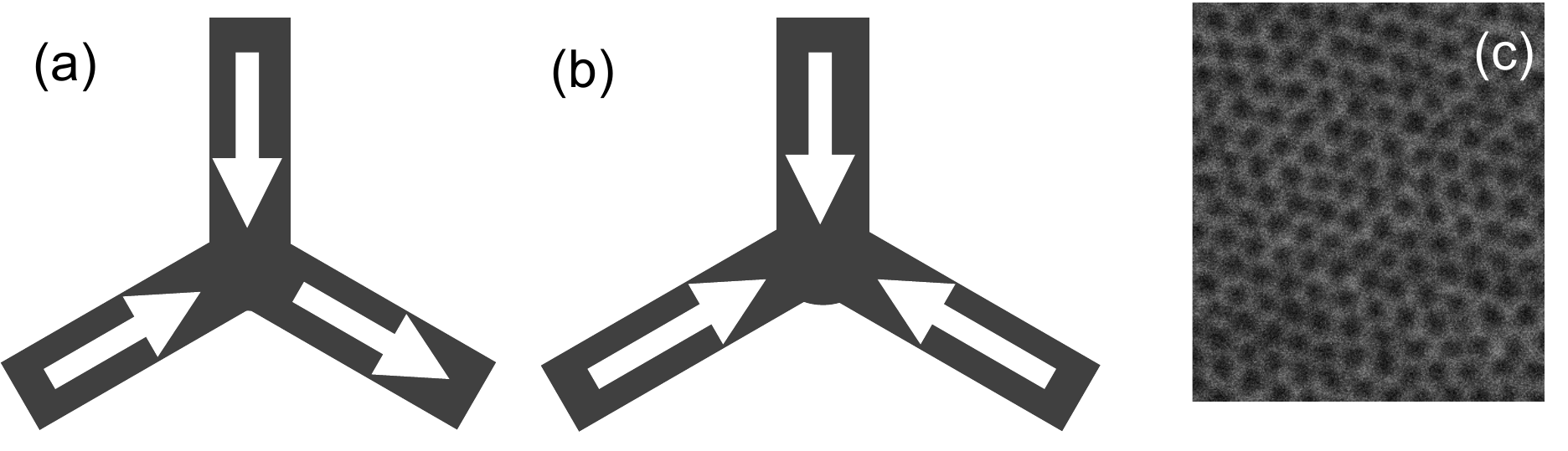} \vspace{-2mm}
\caption {(color online) Moment configuration. (a) Schematic of two-in and 1-out moment configuration on a honeycomb lattice vertex. Here two moments are pointing to the vertex and one point is pointing away from the vertex. (b) All-in moment configuration where all three moments are pointing to the vertex.  (c) The scanning electron micrograph of a typical artificial honeycomb lattice, fabricated using the diblock template synthesis method.
} \vspace{-6mm}
\end{figure}

Previously, researchers have used electron-beam lithography technique to create two-dimensional artificial magnetic honeycomb lattice.\cite{Tanaka,Heyderman,Qi} The typical dimension of the connecting element in this case is of the order of 500 nm (length) $\times$ 50 nm (width) $\times$ 20 nm (thickness). The large typical dimension of individual element (or bond) leads to an effective moment of $\simeq$ 10$^{7}$ Bohr magnetons (for permalloy). Such large magnetic moment results in an exceedingly large inter-elemental dipolar (magnetostatic) energy, $\simeq$ 10$^{4}$ K, in a permalloy honeycomb.\cite{Nisoli} The inter-elemental energy is basically the dipolar interaction energy between two magnetic moments aligned along the honeycomb lattice elements (see Fig. 1a-b). Therefore, the experimental system is well into its ordered state at any reasonable measurement temperature and thermal fluctuations could not induce spin flip or induce the development of a new phase. The inter-elemental energy, $E_D$, is the controlling factor in the temperature dependent magnetic phase transition. The inter-elemental energy, $E_D$, must be smaller than the thermal energy, $E_{th}$ = $k_B$T, at a measurement temperature $T$, so that the thermal fluctuation can overcome this energy barrier. More recently, we proposed a sample fabrication design, which resulted in ultra-small connecting element, with a typical dimension of 12 nm (length) $\times$ 5 nm (width) $\times$ 5 nm (thickness), and the large sample size of the honeycomb lattice. At this element size, the estimated inter-elemental energy is $\simeq$ 12 K, which is small enough to allow temperature to be a feasible tuning parameter to explore the predicted phase diagram in artificial honeycomb lattice. Using small angle neutron scattering and neutron reflectometry measurements on the recently designed artificial honeycomb lattice, we demonstrated the development of temperature dependent magnetic phases, including the spin solid order at low temperature, in this system.\cite{Brock,Artur} 

\begin{figure}
\centering
\includegraphics[width=9 cm]{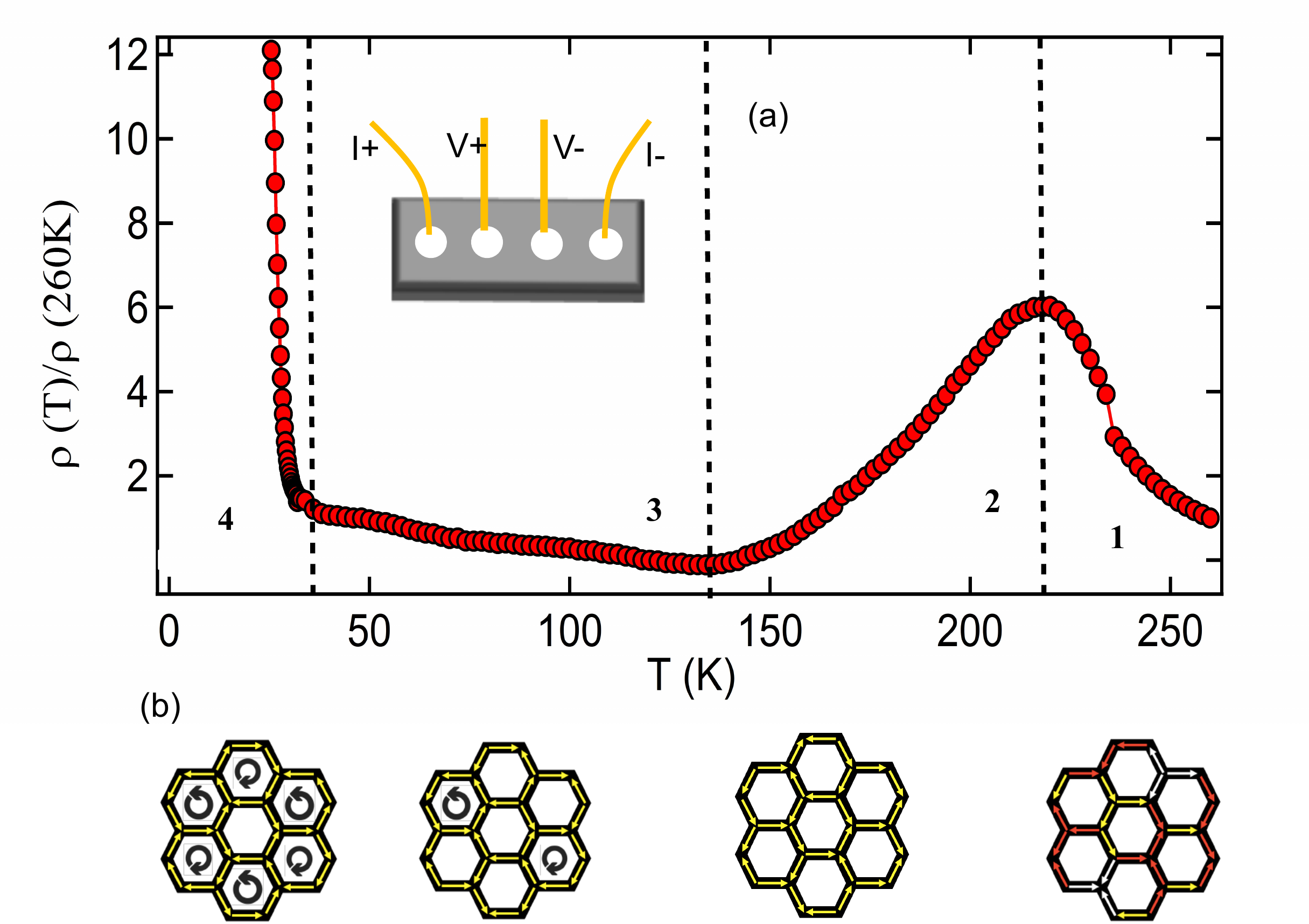} \vspace{-5mm}
\caption {(color online) Electrical resistivity and multiple magnetic regimes in artificial permalloy honeycomb lattice of ultra-small element. (a) Electrical resistivity, $\rho$, as a function of temperature. Different magnetic regimes, as evidenced from the neutron scattering measurements (see text and associated references), are separated by dashed lines. Inset shows the schematic of four probe electrical measurements configuration where gold wires are attached to the sample using silver paint. The average size of silver dot was $\simeq$ 0.5 mm, while the sample size was 4 mm (width) $\times$ 7 mm (length). (b) Schematic depiction of various magnetic phases in the newly designed artificial magnetic honeycomb lattice as measurement temperature is reduced. At low temperature, $T \leq$ 30 K, the system enters into a spin solid order regime.
} \vspace{-6mm}
\end{figure}

The fabrication of two-dimensional permalloy honeycomb lattice of ultra-small element involves the synthesis of diblock copolymer template, coupled with reactive ion etching and material deposition using a state of the art deposition stage in a near parallel configuration.  For this purpose, we utilized a diblock copolymer polystyrene-b-poly-4-vinyl pyridine (PS-b-P4VP) of molecular weight 23 K Dalton with the volume fraction of 70\% PS and 30\% P4VP. The diblock solution in toluene was spin casted at 2500 rpm on top of the silicon substrate to create a thin polymer film. The sample was solvent annealed at 25$^{o}$ C for 12 hours in a mixture of tetrahydrofuran:toluene (80:20 v/v) environment. Subsequent cleaning of the polymer sample in ethanol solvent removes one component of the diblock copolymer (P4VP) and generates a large uniform hexagonal nanoporous template of polystyrene with the pore dimeter of $\simeq$ 12 nm and a lattice spacing of $\simeq$ 30 nm.\cite{Park} Next, the diblock template was used as the etching mask to transfer the hexagonal pattern to the underlying silicon substrate using CF$_{4}$ (carbon tetrafluoride)-based reactive ion etching technique. After reactive ion etching, the polymer template is usually etched away. We cleaned the substrate in toluene solvent to remove any polymer residue. The top layer of the substrate resembled a honeycomb lattice pattern. This topographical property was exploited to create metallic honeycomb lattice by depositing permalloy, Ni$_{0.81}$Fe$_{0.19}$, in near parallel configuration (within 1 degree) using a new sample holder, specifically designed to limit the deposition to the top of the substrate only, in an electron-beam evaporator. The substrate was rotated at a moderate constant speed about its axis during the deposition process to create uniformity in the film thickness. The latter process helps in achieving the two-dimensional character of the structure. The scanning electron micrograph of a typical honeycomb lattice is shown in Fig. 1c. Electrical measurements are performed using the four probe configuration on a 4 mm $\times$ 7 mm (length) size newly synthesized honeycomb lattice sample, shown in the inset of Fig. 2a. Electrical contacts were made using ultra-high purity silver paint and gold wire of 99.99\% purity. Magnetic hysteresis measurement was performed using a commercially available Quantum Design magnetometer with a base temperature of $\simeq$ 2 K.

\begin{figure}
\centering
\includegraphics[width=8.8 cm]{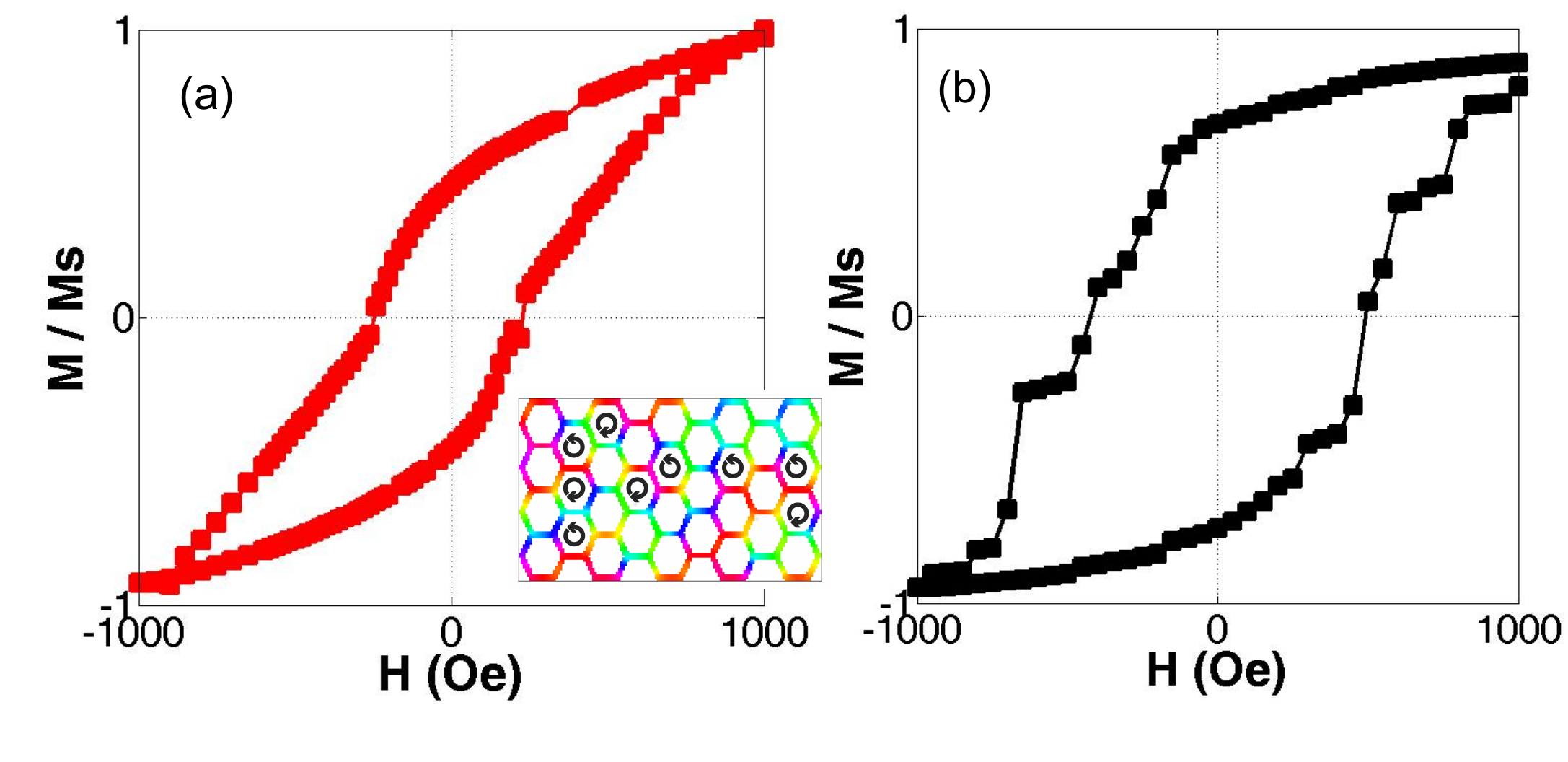} \vspace{-6mm}
\caption{(Color online) Magnetic measurements in applied magnetic field and micromagnetic simulation results of permalloy honeycomb lattice. (a) M vs. H measurement at $T$ = 5 K. Magnetic field was applied in-plane to the sample. (b) Experimental data is compared with micromagnetic (MM) simulation. Near zero field, the artificial honeycomb lattice exhibits pairs of vortex loops of opposite chiralities (shown in the inset of Fig. a). The magnetic hysteresis loop obtained from micromagnetic (MM) simulation, fig. b, agrees well with the experimental data. The MM simulation was performed using 0.2 $\times$ 0.2 nm$^{2}$ mess size, with magnetic field applied in-plane to the lattice. 
} \vspace{-6mm}
\end{figure}

We plot the normalized electrical resistivity as a function of temperature in Fig.2a. There are several different temperature regimes, indicative of different magnetic phases in the system.\cite{Brock,Artur} Previously, detailed small angle neutron scattering (SANS) and reflectometry measurements provided insight into magnetic correlation in different temperature regimes in this system. As described schematically in Fig. 2b, the system undergoes a transition from the paramagnetic phase, where the magnetization across the lattice is described by the random distribution of '2-in \& 1-out' (or, vice-versa) and 'all-in or all-out' moment arrangements on the honeycomb vertices, to the short-range ordered spin ice phase, primarily involving '2-in \& 1-out' (or, vice-versa) configuration only. Here, '2-in \& 1-out' refers to a peculiar configuration of two moments, aligned along the honeycomb lattice elements due to the shape and magnetocrystalline anisotropy, are pointing to the vertex and one moment is pointing away from the vertex in the system. Thus, it obeys the quasi-ice rule. Similarly, all-in and all-out moment configurations indicate situations where all moments are either pointing to the vertex or pointing away from the vertex, respectively. As temperature is reduced further, the magnetization pattern changes to the magnetic charge ordered state, which consists of random pair of vortex loops of opposite chiralities. At much lower temperature, $T \leq$ 30 K, the system is found to develop long range spatial correlation of spin solid order, manifested by the ordered arrangements of the vortex loops of opposite chiralities of magnetic moments. We observe that the electrical resistivity exhibits a very sharp enhancement of more than two orders of magnitude, compared to the value at $T$ = 40 K, in the spin solid regime. The artificial honeycomb lattice exhibits strong insulating characteristic in the long range ordered phase. 

The spin solid phase is manifested by the vortex loops of opposite chiralities. This is also confirmed by magnetic hysteresis measurements at low temperature. In Fig. 3a, we plot $M$ vs $H$ data at $T$ = 5 K. A sharp transition to a near zero magnetization state near the zero field value is observed in the magnetic hysteresis plot. To understand this, we have performed micromagnetic simulations on artificial permalloy honeycomb lattice of similar element size and thickness by utilizing the Landau-Lifshitz-Gilbert equation of magnetization relaxation in a damped medium.\cite{Brown} For the simulation, we have used exchange stiffness $A$ = 1.0$\times$10$^{-11}$\,J/m, uniaxial anisotropy strength $K_1$ = -1.0$\times$10$^{3}$\,J/m$^{3}$ and damping constant $\alpha$ = 0.2, that are also standard parameters for permalloy ferromagnet. The artificial honeycomb lattice was simulated using 0.2 $\times$ 0.2 nm$^{2}$ mess size on the OOMMF (object oriented micromagnetic framework) platform, with magnetic field applied in-plane to the lattice.\cite{OOMMF} As shown in Fig. 3b of the simulated curve and magnetization profile, the magnetic correlation near zero field is found to be dominated by the distribution of the vortex loops of opposite chiralities, as predicted in the spin solid state.

\begin{figure}
\centering
\includegraphics[width=8.7 cm]{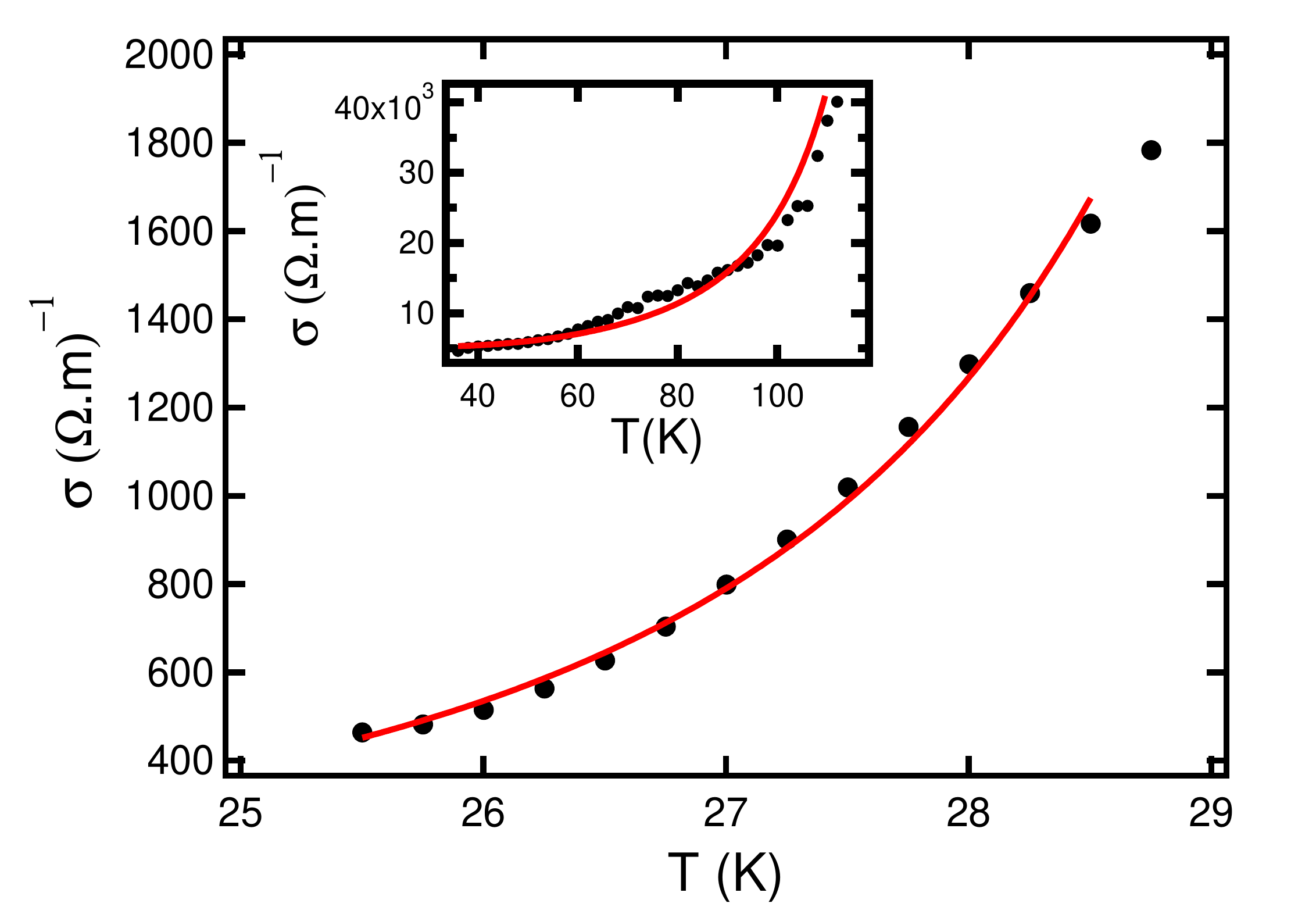} \vspace{-8mm}
\caption{(Color online) Dimensionality analysis of the conductivity data. Electrical conductivity data at low temperature (in the spin solid phase) is fitted using AL model to extract the electrical transport dimensionality of the system. The system manifests a fractional dimensional behavior, $d$ = 0.6(0.04), in the spin solid phase, as opposed to the two-dimensional characteristic at higher temperature. Inset shows the fit to higher temperature conductivity data, which confirms the two-dimensional character of the system ($d$ = 2.06(0.14)).
} \vspace{-6mm}
\end{figure}

The electrical insulating behavior in the spin solid phase at low temperature is unusually strong. This is a very surprising observation. After all, the lattice structure parameters remain unchanged and the only change occurs is in the form of the rearrangement of the magnetization pattern. To understand this, we have performed dimensionality analysis of low temperature electric data using Aslamozov-Larkins (AL) model.\cite{AL} AL model is often used to deduce the electrical dimensionality of a material, especially type-II superconductor where the underlying physics is dominated by the pairs of magnetic vortex and anti-vortex circulations that give rise to additional resistance.\cite{Superconductor,Clippe} Similar magnetic vortices are the building blocks of spin order state. AL model correlates conductivity to a characteristic parameter $t$ vis the following relation: $\sigma$ = ($a + b T$)$^{-1}$ + $c$ $t$$^{-(\epsilon + 1)}$, where $a$, $b$, $c$ are constants and $t$ is given by, $t$ = (1-$T$/$T_0$). Here $T_0$ is the onset to the spin solid phase, $T_0$ = 33 K. The fitting parameter $\epsilon$ is related to the electrical dimensionality $d$ by $d$ = 4 - 2$\epsilon$. We fit the low temperature conductivity data using the AL model, as shown in Fig. 4. The estimated electrical dimensional value is found to be $d$ = 0.6(0.04), which is much smaller than the surface dimensionality (D = 2). It suggests that the electrical properties do not follow the typical surface transport behavior at low temperature. If we use the same model to fit the data in other temperature regimes, then the estimated dimensionality is found to be $d$ = 2.06(0.14), which is close to the full lattice dimensional value of 2 (see the inset in Fig. 4). The two-dimensional character is preserved across the entire temperature range of the measurement, except in the spin solid regime where it exhibits fractional dimensionality.

In Summary, we have observed two interesting electrical properties: first, the system exhibits a strong insulating characteristic as soon as it enters in the spin solid phase and second, the electrical transport follows a fractional dimension at low temperature, which is even smaller than $d$ = 1. In general, a bulk material (of electrical dimensionality three) manifests smaller resistance, compared to the lower dimensional system e.g. two-dimensional system. As the dimensionality reduces, the electrical resistance tends to increase.\cite{Efros} If the honeycomb lattice follows the trend, then the electrical dimensionality of $d$ = 0.6 at low temperature should exhibit a much larger resistance, compared to the full two-dimensional value at higher temperature. The experimental observation is consistent with this qualitative explanation. It also suggests that the system does not follow the surface-like transport at low temperature, which is a common property of two-dimensional systems. However, different lattice structure, for instance square spin ice configuration, may exhibit different electrical dimensionality than the honeycomb structure, as the loop state of spin solid order cannot be generated in the latter case. Interestingly, the surface-like behavior is preserved throughout the measurement, except in the spin solid phase at low temperature, in artificial magnetic honeycomb lattice. The underlying physics behind such a drastic change in the electrical transport pattern is not understood at this point. It can be argued that the ordered arrangement of vortex loops plays an indirect role in this process. However, we do not have any direct evidence, which can corroborate this hypothesis. Future research works, especially theoretical, that can correlate these two phenomena are highly desirable. It will also be helpful in understanding the role of magnetism in the electrical transport properties in two-dimensional lattice.

The research at MU is supported by the U.S. Department of Energy, Office of Basic Energy Sciences under Grant No. DE-SC0014461.

\clearpage


\begin{thebibliography}{99}


\bibitem{Sasha} S. Stankovich, D. Dikin, G. Dommett, K. Kohlhaas, E. Zimney, E. Stach, R. Piner, S. Nguyen and R. Ruoff, \textit{Nature} \textbf{442}, 282 (2006).

\bibitem{Feng} B. Feng, Z. Ding, S. Meng, Y. Yao, X. He, P. Cheng, L. Chen and K. Wu, \textit{Nano Lett.} \textbf{12}, 3507 (2012).

\bibitem{Ataca} C. Ataca and S. Ciraki, \textit{J. Phys. Chem. C} \textbf{115}, 13303 (2011).

\bibitem{Nisoli} C. Nisoli, R. Moessner and P. Schiffer, \textit{Rev. Mod. Phys.} \textbf{85}, 1473 (2013).

\bibitem{Tanaka} M. Tanaka, E. Saitoh, H. Miyajima, T. Yamaoka, and Y. Iye, \textit{Phys. Rev. B} \textbf{73}, 052411 (2006).

\bibitem{Branford} W. R. Branford, S. Ladak, D. E. Read, K. Zeissler, and L. F. Cohen, \textit{Science} \textbf{335}, 1597 (2012).

\bibitem{Moller} G. Moller and R. Moessner, \textit{Phys. Rev. B} \textbf{80}, 140409 (R) (2009).

\bibitem{Chern} G. W. Chern, P. Mellado, and O. Tchernyshyov, \textit{Phys. Rev. Lett.} \textbf{106}, 207202 (2011).

\bibitem{Le} B. Le, D. Rench, R. Misra, L. O'Brien, C. Leighton, N. Samarth and P. Schiffer, \textit{New J. Phys.} \textbf{17}, 023047 (2015).

\bibitem{Brock} B. Summers, Y. Chen, A. Dahal and D. K. Singh, \textit{Sci. Rep.} \textbf{7}, 16080 (2017).

\bibitem{Heyderman} V. Kapaklis, U. Arnalds, A. Farhan, R. Chopdekar, A. Balan, A. Scholl, L. J. Heyderman and B. Hjörvarsson, \textit{Nature Nanotechnology} \textbf{9}, 514 (2014).

\bibitem{Qi} Y. Qi, T. Brintlinger, and J. Cumings, \textit{Phys. Rev. B} \textbf{77}, 094418 (2008).

\bibitem{Artur} A. Glavic, B. Summers, A. Dahal, J. Kline, W. Van Herck, A. Sukhov, A. Ernst and D. K. Singh, \textit{Advanced Science} \textbf{2018}, 1700856; DOI: 10.1002/advs.201700856

\bibitem{Park} S. Park, B. Kim, O. Yavuzcetin, M. Tuominen and T. Russell, \textit{ACS Nano} \textbf{2}, 1363 (2008).

\bibitem{Brown} W. F. Brown, \textit{Micromagnetics}, New York Wiley (1963).

\bibitem{OOMMF} OOMMF software is developed by NIST staff. More detail can be found at: http://math.nist.gov/oommf/

\bibitem{AL} L. G. Aslamozov and A. I. Larkins, \textit{Phys. Lett.} \textbf{26A}, 238 (1968).

\bibitem{Superconductor} R. Ivens, R. Wernhardt, M. Rosenberg, S. Losch and S. Sack, \textit{Sup. Sci. Tech.} \textbf{5}, 16 (1992).

\bibitem{Clippe} P. Clippe, C. Laurent, S. K. Patapis and M. Ausloos, \textit{Phys. Rev. B} \textbf{42}, 8611 (1990).

\bibitem{Efros} A. L. Efros and B. I. Shklovskii, \textit{J. Phys. C: Solid State Phys.} \textbf{8}, L49 (1975).





\end{thebibliography}
\end{document}